# A Survey of Learning Management Systems and Synchronous Distance Education Tools

Khondkar Islam, Pouyan Ahmadi, and Salman Yousaf
George Mason University
kislam2, pahmadi, syousaf3@gmu.edu

*Abstract*— Although compelling assessments have been quite frequently examined in recent years, more studies are required to yield a better understanding of several Distance Learning (DL) methods where Learning Management Systems (LMSs) significantly affect student learning process. Most studies in this area do not consider the effect of varying web-facilitated DL application tools. To address these drawbacks, the objective of our study is to compare two LMSs and four synchronous distance education tools (SDET). The comparisons confirm the superiority of Moodle Integrated Synchrotrons Teaching Conferencing (MIST/C), which seems to be the most practical, convenient and modest distance education tool offered in the market today because it is open source and has a second mirrored whiteboard for simulteaching that is not available with any other system.

## I. Introduction

Distance education (DE) is an effective mode of learning, if the delivery and exchange of education content are facilitated properly. Web-based DE application tools used by students and faculty are key components required to achieve this, by making the learning process easy and effective while eliminating unnecessary difficulty. Some of these tools are difficult both to learn and use, and thus turn out to be an obstacle for faculty and students [1]. A survey of popular DE application tools being used today is presented in this paper to compare their key features, ease-of-use and learning curve. This includes comparisons of two learning management systems (LMSs): one developed by a commercial institution, and the other an open source. Comparisons of four synchronous distance education tools (SDETs) are also included here: one is a commercial product and three are open source.

### A. DE LMS and SDET Requirements

DE LMS and SDET must offer a user-friendly graphical user interface, simple navigation options, and have enhanced security features to deter unauthorized access to the system and files [2]. The system should not be bandwidth intensive, and must be capable of processing audio and video streaming in a distributed network setting. Course creation and management must be easy, and the system must support common file types. There should be an option to reuse course contents so instructors can reuse contents in other sections of the same course or during another semester with minor modifications.

The virtual environment provided by DE LMS and SDET should appear sufficiently real to create an atmosphere where faculty and students feel they are interacting in person. This raises the learning motivation of students and teaching enthusiasm of instructors that promote class participation using the *discussion board* and *chat* room. It is therefore necessary for the LMS and SDET to have easy-to-use and effective interactive communication options such as the discussion board, chat room, email, etc. [2].

Early research suggests web users need to be provided with an effective usable environment because it drives substantial savings and achieves better performance. In academia, effective LMS and SDET need little instructor time to set up and manage the course, while improving the learning experience of students. It is important for the LMS and SDET to be not cluttered with too many appealing usage options, as that can be confusing for students and the instructor. Only features that meet course objectives and are relevant to a sound-learning environment for designing an effective course should be included in the LMS and SDET. Since usability is critical, the LMS and SDET must be easy to use and to learn, and must offer options that are easy to remember. Considering web usability, tools must have web pages that are easy to navigate and must display information in an organized manner so users do not have to struggle to find what they are looking for. Pedagogical usability enables users to learn effectively and retain the skills and knowledge learned; it is integrated with technology usability, which is referred to ease of use and usefulness of the technology [1]. Students do not have a high degree of pedagogical usability when technology usability is poor.

### B. Communication Tools

Zaina et al [2] describe *chat* as a synchronous communication tool that allows students to receive immediate feedback on a subject, which helps to understand group reflection of the subject matter. The researchers view *discussion board* as an asynchronous communication tool where faculty and students post messages to share and debate ideas. One of the positive aspects of the discussion board is that it allows fusion among the group and lets them evaluate and think about a post before responding to it. In addition, the messages are saved and can be re-visited by the class at any time. It can be beneficial if chat conversations are also saved in the event a chat message is needed later.

*C. Grade Book Component*

The *grade book* is a critical component of DE. Since students do not interact with the instructor in person, it is important for the faculty to be transparent with student grades. The grade book feature in most LMSs and SDETs lets the instructor post scores of home assignments, lab work, exams, etc. that the student is able to see, and monitor progress in class. This tool has to be easy to use and easy to understand by the instructor and students [1].

*D. Knowledge Assessment*

Zaina et al. [2] say knowledge assessment is essential in DE and is possible via examinations. The LMS and SDET must offer an effective tool to the instructor for developing exams with a time window to take the exam. This tool must offer statistics of each question that gives a snapshot of how many students answered the question correctly. It enables the instructor to have a better understanding of the areas where students are deficient, and serves as useful information for the instructor to emphasize more on deficient areas to address lacking, and make necessary changes to the course content. A controversial aspect of knowledge assessment is the ability to administer proctored examinations to online students. While some educators see this as a requirement, others do not believe it has been achieved yet in a cost-effective way.

*E. Administrator Role*

Most LMSs and SDETs have an administrator whose task is to publish or set up the course, register users and address technical problems with the system. The responsibilities of the administrator should be restricted to just these tasks, because it would be difficult and time consuming for the instructor and students to depend on the administrator with course related matters, which should be the responsibility of the instructor. It would also overwhelm the administrator with problem tickets and lead to inefficient usage of the LMS and SDET [2].

The next section compares functionalities of Blackboard (BB) [3] and Moodle LMS to determine which one is better.

*F. Blackboard and Moodle*

Unal et al [1] conducted a study on usability of BB and Moodle LMSs. 135 students participated during fall 2008 and spring 2009 semesters. Their study shows Moodle, an open source LMS, was favored by participants over commercial LMS, BB. Only the Discussion Board module of BB fared slightly better than Moodle. Apart from quantitative comparison of participants' responses, the authors analyzed components of both LMSs that students found useful or better than the other.

Blackboard Corporation, founded in 1997, developed BB. It has thousands of deployments in over 60 countries and is available in 8 major languages [1]. Martin Dougiamas founded Moodle in 2001 and has over 70,000 active deployments in 222 countries translated into 75 languages [4]. The Moodle LMS is an open source software package and has a flexible modular design that allows users to select and implement extensions from thousands of available options to design their customized version of Moodle [1].

Unal et al. [1] asked participants about their experiences with Moodle and BB, and provide feedback on the following components:
- Course format and layout of both LMS
- Announcements of BB and News Forum of Moodle
- Course Documents of BB and Lessons of Moodle
- Assignment Manager of BB and Assignment/Activity of Moodle
- Discussion Board of BB and Discussion Forum of Moodle
- Collaboration Tools of both LMS
- Communications of BB and Moodle
- My Grade of both LMS

*Course format and layout* of BB is quite different from Moodle because it has a layout for the instructor that is in standard compartmentalized format and cannot be changed. It has different sections for each tool that has options for the instructor to manage the course, users, and course contents. Moodle on the other hand takes a different approach because it offers the instructor to select from one of the three different formats: weekly, topics and social. Weekly format has activities organized week by week, topics is similar to weekly where each week is referred to a topic. Social format is used as the social forum. There are three columns in the default layout of Moodle. There is one broad column in the middle with two narrow side columns. For this experiment, weekly format was used with course material in the broad column. The study found students favored course format and layout of Moodle over BB.

*Announcements* of BB and *News Forum* of Moodle are the most used modules. In BB, announcements section is on the homepage where posts can be made and seen. In Moodle, News Forum is used for general announcements and is located at the top of the center section. The participants preferred using the News Forum of Moodle over the Announcements module of BB.

The LMS component that is used to deliver course content is critical for an online course. BB has *Course Documents* module to provide course material in text, image and video formats. In Moodle, the *Lessons* module is used for this purpose where a lesson has a series of interactive pages. The student must select an answer to proceed to the next page. Students preferred the Lessons module of Moodle over the Course Documents module of BB.

Discussion is also an important component of LMS. *Discussion Board* of BB is composed of forums where students can select a discussion board by clicking on the name of the board to add new topics or post a reply. Moodle's Discussion Board creates a discussion thread automatically when an instructor creates the forum. Students can reply to the thread and other postings. Students rated the discussion board of BB and Moodle about equal.

*Assignment Submission* feature allows students to upload assignments to the LMS. In this area, the instructor posts assignment with submission link for students to submit assignments by the due date. The survey revealed students favored Moodle over BB for assignment submission.

*Collaboration and group work* is important for an online class to be effective. BB and Moodle have similar elements to allow the instructor to create groups and assign students to individual group manually so they can share documents and send emails to each other, groups or the entire class. A Wiki module is available with both LMSs surveyed and was used by students to work together on a document, and track changes made to the document. Moodle offers an additional feature over BB, which is the option to post profile pictures. This automatically places the student's profile picture where his/her name appears. This feature creates an environment for students to get to know each other in the online environment because the profile picture linked to a profile page with description, location and email address of the student. The study found students preferred the collaboration and communication tools of Moodle over BB.

*Gradebook* module of a LMS or SDET allows instructors to post, update or remove grades of all students in addition to the option to import or export the grade book to an external application. The student can view his/her own grade using this module. BB and Moodle have similar grade book functions providing categorization and statistical reports. The cited comparison found students favoring the grade book of Moodle over that of BB.

This survey clearly shows Moodle to be as effective as BB that can be used as an alternative for online courses. Moodle also offers better technology usability, leading to a greater level of pedagogical usability, and has low total cost of ownership since it is an open source LMS that does not call for licensing expenses of commercial systems. Table 1 outlines the comparison of Moodle and BB.

**Table 1: Comparison of BB and Moodle LMS [1]**

|  | BB | Moodle |
| --- | --- | --- |
| **Format & Layout** | ✗ | ✓ |
| **Announcement** | ✗ | ✓ |
| **Course Docs** | ✗ | ✓ |
| **Assignment Manager** | ✗ | ✓ |
| **Discussion** | ✓ | ✓ |
| **Collaboration** | ✗ | ✓ |
| **Gradebook** | ✗ | ✓ |

*G. Synchronous DE: Network EducationWare (NEW)*

Snow et al. [5] state engineering and technology sectors are dominated by classroom lecture presentation based instructions. This includes lectures by an instructor in the classroom using blackboard and an overhead projector for presentation slides. The smart classroom concept has become popular in recent years because it allows computer generated lecture presentation to be combined with annotations for display to the student audience either in the classroom or to a remote location via the Internet. Using this approach, pre-recorded lectures are used for asynchronous delivery and live classes are made possible for synchronous DE delivery and exchange. It is critical to deliver quality DE material and lectures via the Internet in order to achieve an effective synchronous learning experience. Good quality synchronous DE delivery and exchange can be reached if the students can receive spoken and graphical content without significant delay, are able to ask and respond to questions, and are able to interact with each other during the class period.

NEW is an open source SDET that was developed by computer scientists of the Center of Excellence in Command, Control, Communications, Computing and Intelligence (C4I Center) at George Mason University. It can support both synchronous and asynchronous modes of quality DE content delivery. NEW is beneficial to students and instructors that is not bandwidth intensive in delivering high quality presentation without video because it can do so over 56 kb/s connections. This is made possible because instructor audio is compressed and streamed at 20 kb/s with quality of service (QoS) to guarantee audio delivery with higher priority. Since NEW limits individual page size to 64 KB to ensure low delay, it automatically converts larger slide pages to JPEG images to adhere to this size restriction. It does not require expensive or special hardware platform and complex software, and is easy-to-use and administer. In addition to these positive traits, the application software is entirely open source that allows users to use the source code for education purposes, and freely distribute and use the code in educational and governmental settings. The server side uses MySQL database. Apache web server with PHP scripting language is used by NEW's web portal that provides access to users for DE content. NEW is able to deliver audio graphics materials composed of lecture presentation slides, annotations made on the slides, and presenter's voice to the end clients with a few seconds delay. Without video, approximately 5 MB/h of NEW recordings is required for each class time [5].

Once authenticated by the web server, students use a web portal to access live classes and pre-recorded lectures and slides with NEW. The class is presented by the instructor with a NEW client running on his/her workstation. It is easy-to-use because it is not necessary to learn several controls. Key controls to master are the recorder, whiteboard and floor control. *Recorder* is in a button layout that is used to start a recording and for playbacks. *Whiteboard* looks like a computer drawing tool on which slides are presented.

Instructor can make annotations on the slides that make the learning experience effective and interesting. Annotation graphics are not network capacity intensive because only a few hundred bytes are generated per object. However, the freehand tool generates significant network traffic since few hundred bytes are generated per written character. The NEW client is rate limited, which prevents annotations to interfere with audio. JPEG images, HTML, ASCII text and video can be displayed on the whiteboard [5].

NEW operates in client-server mode where students and instructors run the clients, and the server is responsible for setting up connections, user authentication and content delivery. NEW is inexpensive to set up and operate because the capital outlay is minimal as it needs a basic Linux with Java application server (which can run virtually on MS Windows or Mac OS X) with a 1.8 GHz Pentium III processor, 1 GB memory and a 100 Mb/s network connection to support 40 simultaneous end clients. It is quite simple to install and operate NEW server-side components since most of this is automated. Video function can be used by end clients with 200 kb/s or greater network connectivity. Network level multicasting cannot be used due to its limited deployment over the Internet. NEW uses the open source Transport Layer Multicaster (TLM) that allows it to use TCP to connect client and server behind Network Address Translation (NAT) gateways and firewall systems [5].

OpenSSL, an open source Secure Sockets Layer (SSL) package was added to NEW to handle authentication, and content encryption to meet the security needs of users. NEW may be adapted for use by non-English speaking users since it uses Unicode to support several languages for its display components and controls. The developers have been working to expand NEW's footprint via their on-going effort to port NEW client component suite to Linux/UNIX and Mac OS X operating systems.

*H. Moodle Integrated Synchronous Teaching/Conferencing (MIST/C)*

Pullen et al. [6] have gone beyond NEW to combine asynchronous and synchronous modes to achieve more effective delivery and exchange of DE content to students. They do so by taking advantage of software integration capabilities of a high quality asynchronous DE LMS, Moodle to combine with their existing SDET (NEW) as a basis for the design and implementation of a new synchronous online teaching system called MIST/C.

Like NEW, under control of a master client MIST/C offers audio, video, whiteboard interfaces, floor control, recorder and a playback unit. Similar to NEW, it is not bandwidth intensive because it can operate over a 56 kb/s Internet connection without video and can support video via a better network connection.

Recommendations of George Mason University's Volgenau School of Engineering DE Committee were considered on the features that are necessary in hybrid synchronous/asynchronous online teaching environment to draw upon the functional requirements of MIST/C as outlined in Table 2 [7] below:

**Table 2: MIST/C functional requirements [7]**

| *Customizations* | *Accessible, expandable and improvable* |
|---|---|
| Whiteboard | Able to accept graphic files in real time |
| Authoring formats | PowerPoint, PDF, Keynote, OpenOffice- all participants able to annotate slides during session |
| Video | Common computer formats like mpg, avi, mov and camera |
| Recording of sessions | Automatic on server including chat and be able to render as mpeg for podcasting |
| Interactions | Testing, polling and hand raising, and voice and chat |
| Student Tracking | Login status and participation statistics |
| Configurable to screen | By user and application window capture |
| Breakout | Able to partition class into separate groups |

Changes were made to NEW so that MIST/C runs not only on Windows OS platform, but also on Linux and Mac OS X platforms. *Auto reconnect* feature was added where the Master Client informs the instructor and automatically reconnects to the server during network connection failure, without disrupting face-to-face live class or the recording session. It also automatically uploads the client recording to the server at end of class if approved by the instructor. Another useful feature, *server-side recording and download*, has been made possible where class sessions are automatically recorded on the server, in addition to the client. In the event client-side recording misses a segment of the class session, server will post that missing segment from the server-side recording to Moodle for the students or download it for other use.

The MIST/C development sought to create the simplest possible user interface. Considerable changes were made to the interface in MIST/C over NEW by integrating independent window for each active component such as audio, video, whiteboard, floor control, record control, play control, and master client into a small control window on the screen with toggle buttons to manage components as needed. A second *mirrored* whiteboard window is available for students to see full-size slides on the classroom projector; this is not cluttered with components seen on the master client primary whiteboard window. This feature is a significant advancement for simulteaching, where sets of students in different locations and in the classroom with the instructor are taught simultaneously, and is not available with any other synchronous teaching system.

The MIST/C *whiteboard* is an important component that is used to display static presentation slides and dynamic annotations. NEW supported single-page PDF, JPEG and PostScript formats for the whiteboard, but now MIST/C supports multi-page PDF and sharper PNG slides, and can import any application running on the client machine to the whiteboard. The *floor control* now has a button for the voting interface that can be used by the instructor to post a question, and students can enter their vote in real time. Breakout rooms or groups may now be formed by the instructor using the *Breakout Group Manager* feature by a button on the floor control component so that students of a group may communicate only with members of that group. The instructor is able to either join a particular group to establish two-way communication with group members or maintain supervisory or oversight role to engage in one-way communication with members of all groups [7].

Next section reports the results of comparison made with the commercial product, Elluminate and an open source SDET, Dimdim.

### I. Elluminate vs. Dimdim

Lavolette et al. [8] survey results of Elluminate version 9.0 and Dimdim version 4.5 is presented below. It is worth noting that, Dimdim is an open source SDET. Elluminate was acquired by BB in July 2010 [3] and renamed to Blackboard Collaborate. The researchers collect participant data based on their experience with the interface and features of both systems. Table 3 lists the features of Elluminate and Dimdim.

**Table 3: Features of Elluminate 9.0 and Dimdim 4.5 [8]**

| Features | Elluminate 9.0 | Dimdim 4.5 |
|---|---|---|
| *Communications Tools* | No | No |
| Participants | Unlimited | 20 or less |
| Voice chat | 6 or less | 4 or less |
| Text chat | Yes | Yes |
| Video | 6 or less | 1 |
| *Content Tools* | No | No |
| Guided web browsing | Yes | Yes |
| Interactive whiteboard | Yes | Yes |
| Slide presentation | Yes | Yes |
| Polling and quizzing | Yes | No |
| Multimedia presentation | Yes | No |
| Application sharing | Yes | No |
| Desktop sharing | Yes | Yes (plugin required) |
| Simple feedback | Yes | Yes |
| *Logistics Tools* | No | No |
| Breakout rooms | Yes | No |
| Recording and playback | Yes | Yes |
| Password secured | Yes | Yes |
| Cross platform | Yes | Yes |
| Plugins required | Java | Flash |

The survey had 12 Elluminate participants and 5 Dimdim participants attend a one-hour workshop using Google applications with Elluminate and Dimdim. After the session, they were provided with a set of questions for feedback. The researchers prepared the following five questions for the participants to determine advantages and disadvantages of each system [8]:

- Would you consider using Elluminate/Dimdim in your teaching?
- How easy or difficult was Elluminate/Dimdim to use?
- What was difficult about using Elluminate/Dimdim?
- What do you like about Elluminate/Dimdim?
- Do you have any other comments about Elluminate/Dimdim?

Table 4 lists advantages and disadvantages identified by the researchers in their survey of Elluminate and Dimdim.

**Table 4: Advantages and disadvantages of Elluminate and Dimdim [8]**

| Features | Elluminate | Dimdim |
|---|---|---|
| *Virtual meeting room* | None | None |
| Advantages | Meeting room remains available if presenter logs out or network connection is disrupted. | None |
| Disadvantages | None | Closes meeting if host disconnects or logs out. |
| Advantages | No audio problems encountered; has a wizard to set up audio. | None |
| Disadvantages | None | At times during the workshop, participants experienced audio problems; does not have audio wizard. |
| *Whiteboard* | None | None |
| Advantages | None | Has thumbnail of slides next to the presentation space making it easier for the presenter to navigate slides; presentation slides have good resolution; presenter's mouse pointer automatically appear as laser pointer to other participants when the cursor is in the whiteboard space. |
| Disadvantages | Does not have thumbnails making it difficult to navigate slides; poor resolution of images; presenter must hold down mouse button to make it appear as laser pointer to other participants in whiteboard space. | None |
| *Document Upload* | None | None |
| Advantages | Offers choice of resolution when uploading slides; moderators can upload presentation. | Each presentation is uploaded as a separate document. |
| Disadvantages | Number of simultaneous moderators to upload document is unlimited; adds uploaded slides to the continuous list making it difficult to find start of the presentation that was just uploaded. | Does not offer choice of resolution, just one set resolution; only designated presenter by the host may upload the document. |
| Advantages | None | Only the intended recipient sees the private chat. |
| Disadvantages | Private chat is shared with intended recipient and all moderators; private chat appears in the same window as public chat and difficult to close. | Private chat box blocks part of the whiteboard and cannot be removed; main text-chat window closes when the presentation is changed. |
| Advantages | Easy to locate and use because they are clickable buttons on the interface. | None |
| Disadvantages | None | Difficult to find and use because they are hidden under multiple layers of menus. |

From the results of [8], it is evident that both systems have positive and negative traits and it is entirely up to the user to make the final selection. It is however clear that the participants lean more toward Dimdim since it is free because it is an open source SDET. But since Elluminate has some additional features over Dimdim and because it is a popular commercial product that has been in the market since 2001 and is now incorporated into Blackboard, it is being used by many educational institutions whose participants feel comfortable using it. Dimdim was launched in 2007 and has limited coverage.

*J. Conclusion*

MIST/C and Elluminate have many identical features and fare well in the user community. MIST/C appears to be the most cost-effective, easy-to-use and simple distance education tool available in the market today because it is open source and has a second mirrored whiteboard for simulteaching that is not available with any other system. The comparisons of this chapter validate its rich features and functionalities, which was critical in selecting MIST/C for use by individuals for delivery of DE content to remote users over limited bandwidth networks.